\def\ni{\noindent}
\def\vs{\vskip.3cm}

\def\+{{(+)}}  \def\-{ {(-)} }   \def\0{ {(0)} }
\def\1{ {(1)} }  \def\2{ {(2)} }

\def\sq{Q\kern-6pt/}
\def\sQ{Q\kern-12pt\nearrow}
\documentclass[11pt,eqs]{article}

\usepackage{latexsym}
\usepackage{amssymb}
\def\be{\begin{equation}}             \def\ee{\end{equation}}
\def\ba{\begin{array}{rcl}}           \def\ea{\end{array}}
\def\beqa{\begin{eqnarray} }          \def\eeqa{\end{eqnarray} }
\def\beqalign{\begin{eqalign}}        \def\eeqalign{\end{eqalign}}
             
\def\bsubeq{\begin{subequations}}     \def\esubeq{\end{subequations}}
\def\bitem{\begin{itemize}}           \def\eitem{\end{itemize}}

\def\DJ{\leavevmode\setbox0=\hbox{D}\kern0pt
 \rlap{\kern.04em\raise.188\ht0\hbox{-}}D}
\def\dj{\leavevmode\setbox0=\hbox{d}\kern0pt
 \rlap{\kern.215em\raise.46\ht0\hbox{-}}d}

\newcommand{\bd}{\begin{displaymath}}
\newcommand{\ed}{\end{displaymath}}
\begin{document}

\title{ Dirichlet boundary conditions in type IIB superstring theory and fermionic T-duality
\thanks{This work was supported in part by the Serbian Ministry of Education and Science, under contract No. 171031.}}
\author{B. Nikoli\'c \thanks{e-mail address: bnikolic@ipb.ac.rs} and B. Sazdovi\'c
\thanks{e-mail address: sazdovic@ipb.ac.rs}\\
       {\it Institute of Physics}\\{\it University of Belgrade}\\{\it P.O.Box 57, 11001 Belgrade, Serbia}}
%\date{}
\maketitle

\begin{abstract}

In this article we investigate the relation between consequences of Dirichlet boundary conditions (momenta noncommutativity and parameters of the effective theory) and background fields of fermionic T-dual theory. We impose Dirichlet boundary conditions on the endpoints of the open string propagating in background of type IIB superstring theory with constant background fields. We showed that on the solution of the boundary conditions the momenta become noncommutative, while the coordinates commute. Fermionic
T-duality is also introduced and its relation to noncommutativity is considered. We use compact notation so that type IIB superstring formally gets the form of the bosonic one with Grassman variables. Then momenta noncommutativity parameters are fermionic T-dual fields. The effective theory, the initial theory on
the solution of boundary conditions, is bilinear in the effective coordinates, odd under world-sheet
parity transformation. The effective metric is equal to the initial one and terms with the effective Kalb-Ramond field vanish.

\end{abstract}
\vs

\ni {\it PACS number(s)\/}: 11.10.Nx, 04.20.Fy, 11.10.Ef, 11.25.-w   \par

\section{Introduction}
\setcounter{equation}{0}

In string theory T-duality is an important tool to show the equivalence of different geometries and topologies and in determining some of the genuinely stringy implications on structure of the low energy quantum field theory limit \cite{rabin}. The T-duality rules that relate different curved backgrounds with an Abelian isometry were first constructed by
Buscher \cite{buscher}. These rules are known as the Buscher T-duality rules.

Up to few years ago only T-duality along bosonic coordinates has been considered. Analyzing the gluon scattering amplitudes in $N=4$ super Yang-Mills theory, a new kind of T-dual symmetry, fermionic T-duality, was discovered \cite{ferdual,MW1}. Mathematically, fermionic T-duality is realized within the same procedure as bosonic one, except that dualization is performed along
fermionic directions, $\theta^\alpha$ and $\bar\theta^\alpha$, so it can be considered as a generalization of Buscher T-duality to theories with fermionic variables. The fermionic T-duality maps superstring in certain supersymmetric background to superstring in another supersymmetric background.

On the other hand, in the open string theory besides equations of motion there appear boundary conditions also. In this article we will derive the boundary conditions in pure canonical manner and treat them as canonical constraints as in Refs.\cite{kanonski,BNBS}. Checking consistency of the constraints, it turns out that there are infinite number of consistency conditions. Using Taylor expansion we will rewrite them in compact $\sigma$ dependent form. For the case of Dirichlet boundary conditions all constraints are of the second class and we solve them. In this way we will obtain initial coordinates and momenta in terms of the effective variables.

In the previous articles \cite{bnbsjhep,bnbsnpb} we considered type IIB superstring theory in
pure spinor formulation and applied Busher T-duality along all bosonic directions $x^\mu$. We obtained dual theory and its background fields which represented $N=2$ supermultiplet. This supermultiplet consists of two $N=1$ supermultiplets \cite{bnbsjhep}. The $N=1$ supermultiplet, which is odd under world-sheet parity transformation $\Omega:\sigma\to-\sigma$, in fact contains noncommutativity parameters corresponding to Neumann boundary conditions. Effective theory is initial theory on the solution of boundary conditions with $\Omega$ even coordinates. The other $N=1$ supermultiplet, which is $\Omega$ even, contains background fields of the effective theory.

In Ref.\cite{bnbsjhep} we found that, for the
specific boundary conditions some of the bosonic T-dual background fields
were equal to the noncommutativity parameters and the other ones were equal to the effective background fields. Motivated by this fact, in the
paper \cite{bnbsfd} we found such boundary conditions which produced noncommutativity parameters equal to the fermionic T-dual background fields. In that article we used light-cone canonical analysis in order to find suitable variables (currents) which enable us to perform calculations.

In the present paper we investigate the relation between consequences of Dirichlet boundary conditions (noncommutativity parameters and effective background fields) and background fields of the fermionic T-dual theory. The paper is organized in the following way. First, we introduce
the action of the pure spinor formulation for type IIB superstring
theory keeping quadratic and neglecting ghost terms. We assume that all background fields of type IIB theory: metric tensor $G_{\mu\nu}$, antisymmetric field $B_{\mu\nu}$, two gravitinos $\Psi^\alpha_\mu$ and $\bar\Psi^\alpha_\mu$ and bispinor $F^{\alpha\beta}$ are constant. Also we assume that bispinor $F^{\alpha\beta}$ is invertible and that generalized metric $G_{AB}$ introduced in Section 4 is nonsingular. Then we
perform fermionic T-duality and find the explicit expressions for
T-dual background fields.

From this point we change
approach of Ref.\cite{bnbsfd}. Instead light-cone canonical analysis we rewrite the
action in the compact form in terms of the generalized coordinates $x^A=(x^\mu,\theta^\alpha,\bar\theta^\alpha)$, metric
tensor $G_{AB}$ and Kalb-Ramond field $B_{AB}$. It is unexpected that the components of the generalized metric $G_{AB}$ and Kalb-Ramond field $B_{AB}$ are equal to the fermionic T-dual background fields of the paper \cite{bnbsfd}. Because the action has the same form as in bosonic
case we define the currents analogous with those of Ref.\cite{BNBS} but now in the extended space. It turns out that components of these currents are equal to the
currents obtained by light-cone canonical analysis.

Choosing Dirichlet boundary conditions, $\dot
x^A|_0^\pi=0$, and treating them as canonical constraints, we find the initial coordinates and momenta in terms of the effective
ones. Note that in this case the effective variables are $\Omega$ odd parts of the initial variables. It turns out
that coordinates are commutative, while the momenta are
noncommutative. The reason for noncommutativity is the presence of
the effective coordinates in the expressions for initial momenta.
The noncommutativity parameter is proportional to the generalized
Kalb-Ramond field $B_{AB}$.

We find the form of the initial
theory on the solution of the boundary conditions, which we will call effective
theory. It turns out that the effective generalized coordinates are $\Omega$ odd, while the Lagrangian, which is bilinear in generalized coordinates, is $\Omega$ even. The effective
metric is equal to initial one $G_{AB}$, while the terms with effective
Kalb-Ramond field vanish. 

It is known that noncommutativity parameters and background fields obtained on the solution of Neumann boundary conditions correspond to T-dual background fields \cite{bnbsjhep,bnbsnpb,bnbsfd}, if T-duality is performed along bosonic coordinates. In this paper we will present similar result just considering Dirichlet boundary conditions instead Neumann ones and fermionic T-duality instead bosonic one.

At the end we give some concluding remarks.

\section{Type IIB superstring and fermionic T-duality}
\setcounter{equation}{0}

In this section we will introduce the action of type IIB superstring theory in pure spinor formulation and perform fermionic T-duality.

\subsection{Action}

The action of type IIB superstring theory in pure
spinor formulation up to the quadratic terms \cite{berko,susyNC,BNBSPLB} neglecting ghost
terms as in Ref.\cite{susyNC} is of the form
\begin{eqnarray}\label{eq:SB}
&{}&S(\partial_\pm x, \partial_\pm \theta, \partial_\pm \bar\theta)=\kappa \int_\Sigma d^2\xi \partial_{+}x^\mu
\Pi_{+\mu\nu}\partial_- x^\nu \\&+&\int_\Sigma d^2 \xi \left[
-\pi_\alpha
\partial_-(\theta^\alpha+\Psi^\alpha_\mu
x^\mu)+\partial_+(\bar\theta^{\alpha}+\bar \Psi^{\alpha}_\mu
x^\mu)\bar\pi_{\alpha}+\frac{1}{2\kappa}\pi_\alpha F^{\alpha
\beta}\bar \pi_{\beta}\right ]\, ,\nonumber
\end{eqnarray}
where $G_{\mu\nu}$ is metric tensor, $B_{\mu\nu}$ antisymmetric Kalb-Ramond field, $\Psi^\alpha_\mu$ and $\bar\Psi^\alpha_\mu$ are gravitino fields and bispinor $F^{\alpha\beta}$ is RR field strength. In oreder to simplify calculation we suppose that all background fields are constant. The world sheet $\Sigma$ is parametrized by
$\xi^m=(\xi^0=\tau\, ,\xi^1=\sigma)$ and $\partial_\pm=\partial_\tau\pm\partial_\sigma$. Superspace is spanned by
bosonic coordinates $x^\mu$ ($\mu=0,1,2,\dots,9$) and fermionic
ones, $\theta^\alpha$ and $\bar\theta^{\alpha}$
$(\alpha=1,2,\dots,16)$. The variables $\pi_\alpha$ and $\bar
\pi_{\alpha}$ are canonically conjugated momenta to the coordinates
$\theta^\alpha$ and $\bar\theta^\alpha$, respectively. All spinors are Majorana-Weyl ones and $\Pi_{\pm \mu\nu}=B_{\mu\nu}\pm\frac{1}{2}G_{\mu\nu}$.

On the equations of motion for fermionic momenta $\bar\pi_\alpha$ and $\pi_\alpha$ we have respectively
\begin{equation}\label{eq:impulsi}
\pi_\alpha=-2\kappa \partial_+(\bar\theta^\beta+\bar\Psi^\beta_\mu x^\mu)(F^{-1})_{\beta\alpha}\, ,\quad \bar\pi_\alpha=2\kappa (F^{-1})_{\alpha\beta}\partial_-(\theta^\beta+\Psi^\beta_\mu x^\mu)\, ,
\end{equation}
and the action gets the form
\begin{eqnarray}\label{eq:lcdejstvo}
&{}&S=\kappa \int_\Sigma d^2\xi \partial_+ x^\mu \left[\Pi_{+\mu\nu}+2\bar\Psi^\alpha_\mu(F^{-1})_{\alpha\beta}\Psi^\beta_\nu\right]\partial_-x^\nu \\ &{}& +2\kappa \int_\Sigma d^2\xi \left[ \partial_+\bar\theta^\alpha (F^{-1})_{\alpha\beta}\partial_-\theta^\beta+\partial_+\bar\theta^\alpha (F^{-1})_{\alpha\beta}\Psi^\beta_\nu\partial_-x^\nu+\partial_+x^\mu \bar\Psi^\alpha_\mu (F^{-1})_{\alpha\beta}\partial_-\theta^\beta\right]\, .\nonumber
\end{eqnarray}
Here we assume that RR field strength $F^{\alpha\beta}$ is invertible.

\subsection{Fermionic T-duality}

Because the action has a global shift
symmetry in $\theta^\alpha$ and $\bar\theta^\alpha$ directions, we introduce gauge fields, $(v^\alpha_+, v^\alpha_-)$ and
$(\bar v^\alpha_+, \bar v^\alpha_-)$, to get a local symmetry
\begin{equation}
\partial_\pm\theta^\alpha \to D_\pm \theta^\alpha\equiv\partial_\pm\theta^\alpha+v_\pm^\alpha\, , \quad \partial_\pm\bar\theta^\alpha \to D_\pm
\bar\theta^\alpha\equiv\partial_\pm\bar\theta^\alpha+\bar v_\pm^\alpha\, .
\end{equation}
Also we introduce the Lagrange multipliers
$\vartheta_\alpha$ and $\bar\vartheta_\alpha$ which will impose the
field strengths of gauge fields $v_\pm^\alpha$ and $\bar
v_\pm^\alpha$ to vanish
\begin{equation}
S_{gauge}(\vartheta, \bar\vartheta, v_\pm,\bar v_\pm)=\frac{1}{2}\kappa \int_\Sigma d^2\xi \bar \vartheta_\alpha (\partial_+
v_-^\alpha-\partial_- v^\alpha_+)+\frac{1}{2}\kappa \int_\Sigma d^2\xi  (\partial_+
\bar v_-^\alpha-\partial_- \bar v^\alpha_+)\vartheta_\alpha\, .
\end{equation}
So, we do not introduce new degrees of freedom and the full action is of the form
\begin{equation}\label{eq:auxdejstvo}
{}^\star S(x, \theta,\bar\theta,\vartheta,\bar \vartheta,v_\pm,\bar v_\pm)=S(\partial_\pm x, D_- \theta, D_+ \bar\theta)+S_{gauge}(\vartheta,\bar \vartheta, v_\pm, \bar v_\pm)\, .
\end{equation}

Now we fix $\theta^\alpha$ and $\bar\theta^\alpha$ to zero and obtain the action quadratic in
the fields $v_\pm$ and $\bar v_\pm$
\begin{eqnarray}\label{eq:JED}
&{}&{}^\star S=\kappa \int_\Sigma d^2\xi \partial_+ x^\mu \left[\Pi_{+\mu\nu}+2\bar\Psi^\alpha_\mu(F^{-1})_{\alpha\beta}\Psi^\beta_\nu\right]\partial_-x^\nu \\ &{}& +2\kappa \int_\Sigma \left[ \bar v_+^\alpha (F^{-1})_{\alpha\beta}v_-^\beta+\bar v_+^\alpha (F^{-1})_{\alpha\beta}\Psi^\beta_\nu\partial_-x^\nu+\partial_+x^\mu \bar\Psi^\alpha_\mu (F^{-1})_{\alpha\beta}v_-^\beta\right]\nonumber \\ &{}& +\frac{\kappa}{2} \int_\Sigma d^2\xi \left[\bar \vartheta_\alpha (\partial_+
v_-^\alpha-\partial_- v^\alpha_+)+  (\partial_+
\bar v_-^\alpha-\partial_- \bar v^\alpha_+)\vartheta_\alpha\right]\, .\nonumber
\end{eqnarray}

On the equations of motion for multipliers $\vartheta_\alpha$ and $\bar \vartheta_\alpha$ we obtain $\partial_+
v^\alpha_--\partial_- v^\alpha_+=0$ and $\partial_+
\bar v^\alpha_--\partial_- \bar v^\alpha_+=0$ which gives
\begin{equation}\label{eq:vteta}
\bar v_\pm^\alpha=\partial_\pm \bar\theta^\alpha\, ,\quad v_\pm^\alpha=\partial_\pm \theta^\alpha\, .
\end{equation}
Substituting these expression in (\ref{eq:JED}) we obtain the initial action (\ref{eq:lcdejstvo}).

On the equations of motion for $v_\pm^\alpha$ and $\bar v_\pm^\alpha$ we obtain, respectively
\begin{equation}\label{eq:jed1}
\partial_- \bar \vartheta_\alpha=0\, ,\quad \bar v_+^\alpha=\frac{1}{4}\partial_+ \bar \vartheta_\beta F^{\beta\alpha}-\partial_+ x^\mu \bar\Psi^\alpha_\mu\, ,
\end{equation}
\begin{equation}\label{eq:jed2}
\partial_+ \vartheta_\alpha=0\, ,\quad v_-^\alpha=-\frac{1}{4}F^{\alpha\beta}\partial_-\vartheta_\beta-\Psi^\alpha_\mu \partial_- x^\mu\, .
\end{equation}
Substituting these expression in the action ${}^\star S$
we obtain the dual action
\begin{eqnarray}\label{eq:dualnodej}
&{}& {}^\star S(\partial_\pm x, \partial_- \vartheta, \partial_+
\bar \vartheta)=\kappa\int_\Sigma d^2\xi \partial_+ x^\mu  \Pi_{+\mu\nu}\partial_- x^\nu\, ,\\ &{}&+\frac{\kappa}{8}\int_\Sigma d^2\xi\left[\partial_+\bar \vartheta_\alpha F^{\alpha\beta}\partial_-\vartheta_\beta -4\partial_+x^\mu\bar\Psi^{\alpha}_{\mu} \partial_-\vartheta_\alpha+4\partial_+\bar \vartheta_\alpha\Psi^{\alpha}_{\mu}\partial_- x^\mu\right]\, .\nonumber
\end{eqnarray}
Demanding that dual action has the same form as initial one (\ref{eq:lcdejstvo})
\begin{eqnarray}\label{eq:lcdejstvodual}
&{}&{}^\star S=\kappa \int_\Sigma d^2\xi \partial_+ x^\mu \left[{}^\star \Pi_{+\mu\nu}+2{}^\star\bar\Psi^\alpha_\mu({}^\star    F^{-1})_{\alpha\beta}{}^\star\Psi^\beta_\nu\right]\partial_-x^\nu \\ &{}& +2\kappa \int_\Sigma d^2\xi \left[ \partial_+\bar\vartheta^\alpha ({}^\star F^{-1})_{\alpha\beta}\partial_-\vartheta^\beta+\partial_+\bar\vartheta^\alpha ({}^\star F^{-1})_{\alpha\beta}\Psi^\beta_\nu\partial_-x^\nu+\partial_+x^\mu {}^\star\bar\Psi^\alpha_\mu ({}^\star F^{-1})_{\alpha\beta}\partial_-\vartheta^\beta\right]\, ,\nonumber
\end{eqnarray}
and comparing it with (\ref{eq:lcdejstvo}), we read the dual
background fields (with stars)
\begin{equation}\label{eq:GBdual}
{}^\star B_{\mu\nu}=B_{\mu\nu}+\left[ (\bar\Psi F^{-1}\Psi)_{\mu\nu}-(\bar\Psi F^{-1}\Psi)_{\nu\mu}\right] \, , {}^\star G_{\mu\nu}=G_{\mu\nu}+2\left[ (\bar\Psi F^{-1}\Psi)_{\mu\nu}+(\bar\Psi F^{-1}\Psi)_{\nu\mu}\right]\, ,
\end{equation}
\begin{equation}\label{eq:Psidual}
{}^\star\Psi_{\alpha \mu}=4(F^{-1}\Psi)_{\alpha\mu}\, ,\quad {}^\star\bar\Psi_{\mu\alpha}=-4(\bar\Psi F^{-1})_{\mu\alpha}\, ,
\end{equation}
\begin{equation}\label{eq:Fdual}
{}^\star F_{\alpha\beta}=16(F^{-1})_{\alpha\beta}\, .
\end{equation}
They are well defind because we assume that bispinor $F^{\alpha\beta}$ is invertible. Let us note that two successive dualizations give the initial
background fields.

\section{Canonical analysis of type IIB theory in compact notation}
\setcounter{equation}{0}

We are going to find such noncommutativity parameters corresponding to some boundary conditions which can be related with fermionc T-dual fields (\ref{eq:GBdual})-(\ref{eq:Fdual}).

If we introduce the supercoordinates $x^A=(x^\mu,\theta^\alpha,\bar\theta^\alpha)$ and supermatrices
$$\Pi_{\pm AB}=B_{AB}\pm\frac{1}{2}G_{AB}\, ,$$ as
\begin{equation}
\Pi_{+AB}=\left(
\begin{array}{ccc}
{}^\star\Pi_{+\mu\nu} & -\frac{1}{2}{}^\star\bar\Psi_{\mu\beta} & 0\\
0 & 0 & 0\\
\frac{1}{2}{}^\star\Psi_{\alpha\nu} & \frac{1}{8}{}^\star F_{\alpha\beta} & 0
\end{array}
\right)\, ,\quad \Pi_{-AB}=\left(
\begin{array}{ccc}
{}^\star\Pi_{-\mu\nu} & 0 & \frac{1}{2}({}^\star\Psi^T)_{\mu\beta}\\
-\frac{1}{2}({}^\star\bar\Psi^T)_{\alpha\nu} & 0 & \frac{1}{8}({}^\star F^T)_{\alpha\beta}\\
0 & 0 & 0
\end{array}
\right)\, ,
\end{equation}
then the action (\ref{eq:lcdejstvo}) can be rewritten in the form
\begin{equation}\label{eq:Dejstvo}
S=\kappa\int_\Sigma d^2\xi \partial_+ x^A \Pi_{+AB} \partial_- x^B=-\kappa\int_\Sigma d^2\xi \partial_- x^A \Pi_{-AB} \partial_+ x^B\, .
\end{equation}
From the expression for $\Pi_{\pm AB}$ we read the supersymmetric generalization of the metric, $G_{AB}$, and antisymmetric Kalb-Ramond field, $B_{AB}$,
\begin{equation}\label{eq:GAB}
G_{AB}=\Pi_{+AB}-\Pi_{-AB}=\left(
\begin{array}{ccc}
{}^\star G_{\mu\nu} & -\frac{1}{2}{}^\star \bar\Psi_{\mu\beta} & -\frac{1}{2}({}^\star \Psi^T)_{\mu\beta}\\
\frac{1}{2}({}^\star \bar\Psi^T)_{\alpha\nu} & 0 & -\frac{1}{8}({}^\star F^T)_{\alpha\beta}\\
\frac{1}{2}{}^\star\Psi_{\alpha\nu} & \frac{1}{8}{}^\star F_{\alpha\beta} & 0
\end{array}
\right)\, ,
\end{equation}
\begin{equation}\label{eq:BAB}
B_{AB}=\frac{1}{2}(\Pi_{+AB}+\Pi_{-AB})=\left(
\begin{array}{ccc}
{}^\star B_{\mu\nu} & -\frac{1}{4}{}^\star \bar\Psi_{\mu\beta} & \frac{1}{4}({}^\star \Psi^T)_{\mu\beta}\\
-\frac{1}{4}({}^\star \bar\Psi^T)_{\alpha\nu} & 0 & \frac{1}{16}({}^\star F^T)_{\alpha\beta}\\
\frac{1}{4}{}^\star\Psi_{\alpha\nu} & \frac{1}{16}{}^\star F_{\alpha\beta} & 0
\end{array}
\right)\, .
\end{equation}
Note that $G_{BA}=(-)^{A+B+AB}G_{AB}$ and $B_{BA}=-(-)^{A+B+AB}B_{AB}$. Consequently, we have $x^A G_{AB}y^B=y^A G_{AB}x^B$ and $x^A B_{AB} x^B=0$.

The momenta canonically conjugated to the coordinates $x^A$ are
\begin{equation}\label{eq:impulsA}
\pi_A=\frac{\partial_L \mathcal L}{\partial \dot x^A}=\kappa(G_{AB}\dot x^B-2B_{AB}x'^B)\, .
\end{equation}
The basic Poisson algebra is of the form
\begin{equation}
\left\lbrace x^A(\sigma),\pi_B(\bar\sigma) \right\rbrace=(-)^A \delta^A{}_B \delta(\sigma-\bar\sigma)\, ,
\end{equation}
where $\delta^A{}_B$ is unity operator in the superspace
\begin{equation}
\delta^A{}_B=\left(
\begin{array}{ccc}
\delta^\mu{}_\nu & 0 & 0 \\
0 & \delta^\alpha{}_\gamma & 0\\
0 & 0 & \delta^\beta{}_\delta
\end{array}
\right)\, .
\end{equation}
Let us now introduce the currents
\begin{equation}\label{eq:gdstruja}
J_{\pm A}=\pi_A+2\kappa\Pi_{\pm AB}x'^B\, ,\quad J^A_\pm\equiv (G^{-1})^{AB}J_{\pm B}\, ,
\end{equation}
which satisfy Abelian Kac-Moody algebra
\begin{equation}\label{eq:algebraJ}
\left\lbrace J_{\pm A}(\sigma), J_{\pm B}(\bar\sigma)\right\rbrace=\pm 2\kappa G^{st}_{AB}\delta'\, ,\quad \left\lbrace J_{\pm A}(\sigma), J_{\mp B}(\bar\sigma)\right\rbrace =0\, .
\end{equation}
Here we introduced supertransposition
\begin{equation}
X_{AB}=\left(\begin{array}{cc}
A & B\\
C & D
\end{array}
\right)\, ,\quad X^{st}_{AB}=(-)^{A+AB}X_{BA}=\left(\begin{array}{cc}
A^T & C^T\\
-B^T & D^T
\end{array}
\right)\, ,
\end{equation}
where $T$ is related to the ordinary transposition. Note that $G_{AB}^{st}=(-)^B G_{AB}$.

Using the expression for canonical momenta the currents get the form
\begin{equation}\label{eq:Astruja}
J_{\pm A}=\kappa G_{AB}\partial_{\pm} x^B\, ,
\end{equation}
from which we can express $\tau$ and $\sigma$ derivative of $x^A$ in terms of the currents
\begin{equation}\label{eq:izvodi}
\dot x^A=\frac{(G^{-1})^{AB}}{2\kappa}(J_{+B}+J_{-B})\, ,\quad x'^A=\frac{(G^{-1})^{AB}}{2\kappa}(J_{+B}-J_{-B})\, .
\end{equation}
From the definition of the canonical Hamiltonian
$$\mathcal H_c=\dot x^A\pi_A-\mathcal L\, ,$$
using the expression (\ref{eq:izvodi}), (\ref{eq:impulsA}) and (\ref{eq:Dejstvo}), we obtain canonical Hamiltonian in terms of the currents 
\begin{equation}\label{eq:noviHam}
H_c=\int d\sigma \mathcal H_c\, ,\quad \mathcal H_c=T_--T_+\, ,\quad T_{\pm}=\mp\frac{1}{4\kappa}J^A_\pm G_{AB} J^B_\pm\, .
\end{equation}
Using the definition of the current with upper index (\ref{eq:gdstruja}) we rewrite the energy-momtum tensor components in the form
\begin{equation}\label{eq:tei}
T_{\pm}=\mp\frac{1}{4\kappa}J_{\pm A}[(G_{st})^{-1}]^{AB}J_{\pm B}\, ,
\end{equation}
Let us stress that $[(G_{st})^{-1}]^{AB}=(-)^{A+B}[(G^{-1})_{st}]^{AB}$. 

The energy-momentum tensor components are in Sugawara form and satisfy two independent Virasoro algebras
\begin{equation}
\left\lbrace T_\pm(\sigma),T_\pm(\bar\sigma)\right\rbrace=-\left[T_{\pm}(\sigma)+T_\pm(\bar\sigma)\right]\delta'\, ,\quad  \left\lbrace T_\pm(\sigma),T_\mp(\bar\sigma)\right\rbrace=0\, .
\end{equation}
For further analysis it is useful the following relation
\begin{equation}\label{eq:HJ}
\left\lbrace H_c, J_{\pm A}\right\rbrace=\mp J'_{\pm A}\, .
\end{equation}

\section{Boundary conditions and fermionic T-duality}
\setcounter{equation}{0}

We are looking for such boundary conditions that corresponding noncommutativity parameters are just the background fields of the fermionic T-dual theory (\ref{eq:GBdual})-(\ref{eq:Fdual}).

Varying the Hamiltonian (\ref{eq:noviHam}) we obtain
\begin{equation}\label{eq:korisno2}
\delta H_c=\delta H_c^{(R)}-\tilde\gamma_A^{(0)}\delta x^A|_0^\pi\, ,
\end{equation}
where $\delta H_c^{(R)}$ has a form $\delta H_c^{(R)}=C_A \delta x^A+D^A \delta \pi_A$. It contains variations $\delta x^A$ and $\delta\pi_A$ but does not contain corresponding variations of the $\sigma$ derivatives $\delta x'^A$ and $\delta \pi'_A$. The boundary term was obtained, after partial integration, from parts of the form $E_A\delta x'^A$ and it yields
\begin{equation}
\tilde\gamma_A^{(0)}=\Pi_{+AB} J^B_-+\Pi_{-AB} J^B_+\, .
\end{equation}
Because the Hamiltonian is time translation generator it must have well defined functional derivatives with respect to the coordinates and momenta. Consequently, boundary term must vanish
\begin{equation}\label{eq:GU}
\tilde\gamma_A^{(0)}\delta x^A|_0^\pi=0\, .
\end{equation}
We choose Dirichlet boundary conditions
\begin{equation}\label{eq:gu1}
\gamma_A^{(0)}|_0^\pi=0\, ,\quad \gamma_A^{(0)}=2\kappa G_{AB}\dot
x^B=J_{+A}+J_{-A}\, .
\end{equation}
The fact that velocity $\dot x^A$ is zero at string endpoints means that the endpoints do not move. Consequently, string endpoints are fixed, $\delta x^A|_0^\pi=0$, and they solve boundary conditions (\ref{eq:GU}). Applying Dirac consistency procedure we obtain infinite set of the constraints
\begin{equation}
\gamma^{(n)}_A|_0^\pi=0\, ,\quad \gamma_A^{(n)}=\left\lbrace H_c, \gamma_A^{(n-1)} \right\rbrace\, .(n=1,2,3,\dots)
\end{equation}
With the help of the relation (\ref{eq:HJ}) using Taylor expansion
\begin{equation}
\Gamma_A(\sigma)=\sum_{n=0}^\infty \frac{\sigma^n}{n!}\gamma_A^{(n)}|_0\, ,
\end{equation}
we rewrite these infinite sets of consistency conditions at $\sigma=0$ in compact, $\sigma$ dependent form
\begin{equation}
\Gamma_A(\sigma)=J_{+A}(-\sigma)+J_{-A}(\sigma)\, .
\end{equation}
In the similar way we can write the consistency conditions at $\sigma=\pi$. If we impose $2\pi$ periodicity of the canonical variables, the solution of the constraints at $\sigma=0$ also solve the constraints at $\sigma=\pi$.

Using the algebra of the currents (\ref{eq:algebraJ}) we obtain the algebra of the constraints
\begin{equation}
\left\lbrace \Gamma_A(\sigma),\Gamma_B(\bar\sigma) \right\rbrace=-4\kappa G^{st}_{AB}\delta'\, .
\end{equation}
Because we assume that metric $G_{AB}$ is nonsingular, the constraints are of the second class.

\section{Solution of the boundary conditions and noncommutativity}
\setcounter{equation}{0}

Appearance of the second class constraints in the theory means that we have either to introduce Dirac brackets or to solve the constraints. The result will not depend on the choice, but for the practical reasons, we will solve the constraints.

Solving the constraint equations
\begin{equation}
\Gamma_A(\sigma)=0\, ,
\end{equation}
we obtain initial variables in terms of the effective ones
\begin{equation}\label{eq:x}
x^A(\sigma)=\tilde q^A(\sigma)\, ,\quad \pi_A=\tilde p_A-2\kappa B_{AB}\tilde q'^B\, .
\end{equation}
Here we introduced new variables, symmetric and antisymmetric
under world-sheet parity transformation $\Omega:\sigma\to -\sigma$. For bosonic variables we use standard
notation \cite{BNBS}
\begin{eqnarray}\label{eq:bv1}
q^A(\sigma)&=&P_s x^A(\sigma)\, ,\quad \tilde
q^A(\sigma)=P_a x^A(\sigma)\, ,\nonumber\\ p_A(\sigma)&=&P_s \pi_A(\sigma)\, ,\quad \tilde
p_A(\sigma)=P_a \pi_A(\sigma)\, ,
\end{eqnarray}
where $P_s$ and $P_a$ are projectors on the $\Omega$ even and odd parts, respectively,
\begin{equation}
P_s=\frac{1}{2}(1+\Omega)\, ,\quad P_s=\frac{1}{2}(1-\Omega)\, .
\end{equation}
Note that all effective independent variables are $\Omega$ odd. Also the initial momenta (not the coordinates) are linear combination of the effective momenta and effective coordinates.

From basic graded Poisson bracket
\begin{equation}
\{x^A(\sigma), \pi_B(\bar\sigma)\}=(-)^A\delta^A{}_B
\delta(\sigma-\bar\sigma)\, ,
\end{equation}
we obtain the corresponding one in $\Omega$ antisymmetric subspace
\begin{equation}\label{eq:pz0}
\{\tilde q^A(\sigma)\, ,\tilde p_B(\bar\sigma)\}=2(-)^A\delta^A{}_B
\delta_a(\sigma\, ,\bar\sigma)\, ,
\end{equation}
where
\begin{equation}
\delta_a(\sigma,\bar\sigma)=\frac{1}{2}\left[
\delta(\sigma-\bar\sigma)-\delta(\sigma+\bar\sigma)\right]\, ,
\end{equation}
is antisymmetric delta function.
Therefore, the momenta $\tilde p_A$ are
canonically conjugated to the coordinates $\tilde q^A$.

We conclude that all supercoordinates are commutative, while the Poisson brackets of momenta are nonzero
\begin{equation}
\left\lbrace \pi_A(\sigma), \pi_B(\bar\sigma)\right\rbrace = 4\kappa B^{st}_{AB}\partial_\sigma \delta(\sigma+\bar\sigma)\, .
\end{equation}
In fact variables $\tilde p$ are $\Omega$ odd parts of the momenta density. Consequently, if we define the momenta as
\begin{equation}
\tilde P_A(\sigma)=\int_0^\sigma d\sigma_1 \tilde p_A(\sigma_1)\, ,
\end{equation}
the noncommutativity relations get the form
\begin{equation}
\left\lbrace \tilde P_A(\sigma), \tilde P_B(\bar\sigma)\right\rbrace = 4\kappa B^{st}_{AB}\theta(\sigma+\bar\sigma)\, ,
\end{equation}
where
\begin{equation}\label{eq:fdelt}
\theta(x)=\left\{\begin{array}{ll}
0 & \textrm{if $x=0$}\\
1/2 & \textrm{if $0<x<2\pi$}\, .\\
1 & \textrm{if $x=2\pi$} \end{array}\right .
\end{equation}
Therefore, the background fields of the fermionic T-dual theory (\ref{eq:GBdual})-(\ref{eq:Fdual}) (except ${}^\star G_{\mu\nu}$) are noncommutativity parameters for boundary conditions (\ref{eq:gu1}).

\section{Effective theory}
\setcounter{equation}{0}

Effective theory is the initial theory on the solution of boundary conditions. In this section we will find the expressions for the Hamiltonian and Lagrangian of the effective theory.

Because the Hamiltonian is bilinear expression in currents, first we have to find the currents $J_{\pm A}$ on the solution of boundary conditions. Substituting the solution (\ref{eq:x}) in the expressions for the currents $J_{\pm A}$ (\ref{eq:gdstruja}) and energy-tensor components (\ref{eq:tei}), we obtain
\begin{equation}
J^{eff}_{\pm A}\equiv J_{\pm A}|_{\Gamma=0}=\tilde p_A\pm \kappa G_{AB}\tilde q'^B\, ,
\end{equation}
\begin{equation}
T^{eff}_{\pm}\equiv T_{\pm}|_{\Gamma=0}=\mp\frac{1}{4\kappa}J^{eff}_{\pm A}[(G_{st})^{-1}]^{AB}J^{eff}_{\pm B}\, .
\end{equation}
Comparing these relations with (\ref{eq:gdstruja}) and (\ref{eq:tei}), we find the transition rule from the initial to the effective theory
\begin{equation}\label{eq:trule}
x^A\to\tilde q^A\, ,\quad \pi_A\to \tilde p_A\, ,\quad G_{AB} \to G_{AB}\, ,\quad B_{AB}\to 0\, .
\end{equation}
Let us stress that this is only prescription which enables us to obtain effective from initial theory. Actually, $B_{AB}$ does not vanish but the term in the action with effective antisymmetric background field, $\int d^2\xi \partial_+ \tilde q^A B^{eff}_{AB}\partial_- \tilde q^B$, disappears.

Using the expression for effective Hamiltonian
\begin{equation}
\mathcal H_c^{eff}=T^{eff}_--T^{eff}_+=\frac{1}{2\kappa}\tilde p_A [(G_{st})^{-1}]^{AB} \tilde p_B+\frac{\kappa}{2}\tilde q'^A G_{AB}\tilde q'^B\, .
\end{equation}
we can find effective Lagrangian
\begin{equation}
\mathcal L^{eff}=\dot {\tilde q}^A \tilde p_A-\mathcal H_c^{eff}\, .
\end{equation}
On the equation of motion for momenta $\tilde p_A$ we have
\begin{equation}
\dot {\tilde q}^A=\frac{1}{\kappa}(G^{-1})^{AB}\tilde p_B \, ,
\end{equation}
which gives
\begin{equation}
\tilde p_A=\kappa G_{AB}\dot {\tilde q}^B \, .
\end{equation}
Substituting this relation in the expression for effective Lagrangian we finally obtain
\begin{eqnarray}
&{}&\mathcal L^{eff}=\frac{\kappa}{2}\partial_+ \tilde q^A G_{AB} \partial_- \tilde q^B=\\ &=&\frac{\kappa}{2}\left(\partial \tilde q^\mu {}^\star G_{\mu\nu} \partial_- \tilde q^\nu+\frac{1}{4}\partial_+ \bar\theta_a^\alpha {}^\star F_{\alpha\beta}\partial_- \theta_a^\beta+\partial_+ \bar\theta_a^\alpha {}^\star \Psi_{\alpha\mu}\partial_- \tilde q^\mu-\partial_+ \tilde q^\mu {}^\star \bar\Psi_{\mu\alpha}\partial_-\theta^\alpha_a\right)\, ,\nonumber
\end{eqnarray}
where index $a$ in $\theta^\alpha_a$ and $\bar\theta^\alpha_a$ in according with the definition (\ref{eq:bv1}) means antisymmetrization under world-sheet parity transformation $\Omega$.

The form of the effective Lagrangian confirms transition rule (\ref{eq:trule}).
The effective theory is $\Omega$ even. The result formally has the same form as in pure bosonic case \cite{BNBS} up to the change of indices $A\to\mu$ and take care of order of anticommuting variables. In component notation it means
\begin{eqnarray}
&{}& x^\mu\to \tilde q^\mu\, ,\quad \theta^\alpha\to \theta^\alpha_a\, ,\quad \bar\theta^\alpha\to \bar\theta^\alpha_a\, ,\nonumber\\ &{}& G_{\mu\nu}\to G_{\mu\nu} \, ,\quad \Psi^\alpha_\mu\to \Psi^\alpha_\mu\, ,\quad \bar\Psi^\alpha_\mu\to \bar\Psi^\alpha_\mu\, ,\quad F^{\alpha\beta}\to F^{\alpha\beta}\, ,
\end{eqnarray}
while $B_{\mu\nu}\to 0$.

\section{Concluding remarks}
\setcounter{equation}{0}

In the present article we found some unexpected relation between quite different approaches. On the one side, we investigated influence of the Dirichlet boundary conditions on the endpoints of the open string moving in type IIB superstring background and obtained the noncommutativity properties and effective theory. On the other side, we constructed the fermionic T-dual theory and found explicit expressions for T-dual background fields. Finally, we established the relation of the noncommutativity parameters to fermionic T dual fields. We used the pure spinor formulation of the theory keeping all terms up to the quadratic ones and neglecting ghost terms in the action.

Then we performed fermionic T-duality in the way described in Refs.\cite{ferdual,bnbsfd}. Comparing initial and T-dual theory, we found the expressions for fermionic T-dual background fields. 

On the other side we performed canonical analysis of the theory. To simplify calculations we introduced the extended space-time coordinates $x^A=(x^\mu,\theta^\alpha,\bar\theta^\alpha)$, generalized metric $G_{AB}$ and Kalb-Ramond field $B_{AB}$, and rewrote the action in the form of bosonic string theory. We introduced the currents $J_{\pm A}$ in the extended space-time following the analogy with bosonic string theory. It turned out that they are equal to the currents obtained by light-cone analysis from Ref.\cite{bnbsfd}.
Varying the canonical Hamiltonian and demanding that it has well defined functional derivatives with respect to the coordinates and momenta, we obtained boundary term. We chose Dirichlet boundary conditions, $\dot x^A|_0^\pi\sim (J_{+A}+J_{-A})|_0^\pi=0$ and checked their consistency. We found infinite number of constraints at both string endpoints and introduced one $\sigma$ dependent constraint at each endpoint. The algebra of the $\sigma$ dependent boundary conditions closes on the generalized metric $G_{AB}$. Because it is nonsingular, all constraints are of the second class. Solving the constraints we found that Poisson brackets of the momenta are nonzero, while the Poisson brackets of the coordinates are zero. The noncommutativity parameter is generalized Kalb-Ramond field $B_{AB}$. Taking into account that componets of the the $B_{AB}$ are fermionic T-dual fields (except dual metric ${}^\star G_{\mu\nu}$), we establish relation between fermionic T-duality and noncommutativity.

Plugging the solution of the constraints in the initial theory we obtained effective theory. The effective Lagrangian, bilinear in the effective coordinates, is $\Omega$ even. So, term with generalized Kalb-Ramond field (or all $\Omega$ odd terms) are absent from the effective Lagrangian. Only the terms with generalized metric tensor $G_{AB}$ (\ref{eq:GAB}) ($\Omega$ even terms) survive. Note that components of generalized metric of the initial theory are just fermionic T-dual background fields. Therefore, the effective background fields (except $B_{\mu\nu}$) are equal to the initial background fields and there are no bilinear corrections with $\Omega$ odd fields as in the previous cases with Neumann boundary conditions \cite{bnbsnpb,BNBSPLB}. Only Kalb-Ramond field $B_{\mu\nu}$ is projected out from the effective theory. Let us stress some differences of the present case obtained on the solution of the Dirichlet boundary conditions and all previous cases (bosonic and type IIB superstring) obtained on the solution of Neumann boundary conditions. In all previous cases effective variables, $q^A$ and $p_A$, are $\Omega$ even, while in the present case they are $\Omega$ odd, $\tilde q^A$ and $\tilde p_A$. Also, here we obtained the momenta noncommutativity instead of standard coordinate noncommutativity.

We found how boundary conditions determine type of T-duality and noncommutative variables. In great majority of papers Neumann boundary conditions are connected with bosonic T-duality and coordinate noncommuattivity. In the present paper Dirichlet boundary conditions produce fermionic T-duality and momenta noncommutativity. It gives some interesting consequences. For example, the dual RR field strength ${}^\star F_{\alpha\beta}$ is inverse of the initial one $F^{\alpha\beta}$. In bosonic T-duality similar relation holds for the corresponding metric tensors ${}^\star G^{\mu\nu}$ and $G_{\mu\nu}$. Therefore, the noncommutativity parameter of fermionic momenta $\Theta_{\alpha\beta}$, which is proportional to ${}^\star F_{\alpha\beta}$, is inverse of background field $F^{\alpha\beta}$ in fermionic part of the dual action $\partial_+\bar\vartheta_\alpha F^{\alpha\beta}\partial_-\vartheta_\beta$. On the other hand, momenta noncommutativity means that product of uncertainties in two momenta related by nontrivial noncommutativity relation can not be arbitrary small. So, this gives condition in the infrared region instead of the coordinate noncommutativity which gives conditions in ultraviolet region. Consequently, it could affect the long distance physics.

\end{document}